\documentclass[aps,prd,twocolumn,nofootinbib,superscriptaddress]{revtex4-2}
\usepackage{graphicx} \graphicspath{{figures/}}
\usepackage{physics}
\usepackage{amssymb}
\usepackage{amsmath}
\usepackage{amsfonts}
\usepackage{siunitx}
\usepackage[normalem]{ulem}
\DeclareSIUnit{\solarmass}{M_\odot}
\DeclareSIUnit{\parsec}{pc}
\usepackage[lined, vlined, ruled]{algorithm2e}
\SetKwHangingKw{KwHResult}{Result\(\rightarrow\)}
\SetKwFor{PForEach}{parallelly foreach}{do}{end}
\usepackage{tabularx}
\usepackage{hyperref}
\usepackage[capitalize]{cleveref}
\crefname{paragraph}{paragraph}{paragraphs}
\Crefname{paragraph}{Paragraph}{Paragraphs}
\usepackage{xcolor}

\begin{document}

\title{Modular global-fit pipeline for LISA data analysis}

\author{Senwen Deng}
\email{deng@apc.in2p3.fr}
\affiliation{Universit\'e Paris Cit\'e, CNRS, Astroparticule et Cosmologie, F-75013 Paris, France}

\author{Stanislav Babak}
\email{stas@apc.in2p3.fr}
\affiliation{Universit\'e Paris Cit\'e, CNRS, Astroparticule et Cosmologie, F-75013 Paris, France}

\author{Maude Le Jeune}
\email{lejeune@apc.in2p3.fr}
\affiliation{Universit\'e Paris Cit\'e, CNRS, Astroparticule et Cosmologie, F-75013 Paris, France}

\author{Sylvain Marsat}
\email{sylvain.marsat@l2it.in2p3.fr}
\affiliation{Laboratoire des 2 Infinis - Toulouse (L2IT-IN2P3),
	Universit\'e de Toulouse, CNRS, UPS, F-31062 Toulouse Cedex 9, France}

\author{\'Eric Plagnol}
\email{plagnol@apc.in2p3.fr}
\affiliation{Universit\'e Paris Cit\'e, CNRS, Astroparticule et Cosmologie, F-75013 Paris, France}

\author{Andrea Sartirana}
\email{sartirana@apc.in2p3.fr}
\affiliation{Universit\'e Paris Cit\'e, CNRS, Astroparticule et Cosmologie, F-75013 Paris, France}

\date{\today}
\begin{abstract}
	We anticipate that the data acquired by the Laser Interferometer Space
	Antenna (LISA) will be dominated by the gravitational wave signals from
	several astrophysical populations. The analysis of these data is a new
	challenge and is the main focus of this paper. Numerous gravitational wave
	signals overlap in the time and/or frequency domain, and the possible correlation
	between them has to be taken into account during their detection and
	characterization. In this work, we present a method to address the LISA
	data analysis challenge; it is flexible and scalable
	for a number of sources and across several populations. Its performance is demonstrated
	on the simulated data LDC2a.
\end{abstract}

\maketitle
\section{Introduction}\label{sec:introduction}
The Laser Interferometer Space Antenna (LISA) is an ESA-led space-based
gravitational wave (GW) observatory to be launched in 2035. It will operate in
the millihertz frequency range, which is inaccessible to ground-based
detectors.  LISA data analysis is challenging because of the large number of
signals that overlap in time and frequency. We will observe GW signal from (i)
the population of merging massive black hole binaries (MBHBs), these will be
the loudest GW sources; (ii) the population of inspiralling Galactic (and
extra-Galactic) white dwarf binaries (GBs), from which we expect tens of
thousands of individually resolvable sources, making them the most numerous type
of sources in LISA; (iii) the population of inspiralling solar mass black
holes; (iv) the population of extreme mass ratio inspirals. Besides those
populations of binaries, we also expect stochastic GW signal from the energetic
processes in the early Universe (for a complete review, please see
\cite{colpi_lisa_2024}). We anticipate that the data will be GW dominated,
meaning that we do not have a time segment without a detectable GW signal, and
GWs stand above instrumental noise from \SIrange{0.1}{20}{\milli \hertz} (or
maybe even above). This poses a great challenge for data analysis: how to detect,
separate, and correctly characterize all sources. The success of the data
analysis is the key to extracting all the exciting science that we want to do
post-detection.

We are gradually developing the method, implementing it and testing it on the
simulated LISA datasets, which are increasing in complexity and realism. This
paper presents a modular prototype of a global fit pipeline for LISA data
analysis. As a proof of concept, we apply the pipeline to the training dataset
LISA Data Challenge 2 (LDC-2a, a.k.a.\ ``Sangria'')
\cite{le_jeune_2022_7132178}. The simulated data were released to the scientific
community with the deadline for submitting the solution. We are using the
training dataset where all GW sources, their parameters, and the characteristics of
the LISA instrument are revealed. Knowledge of the true parameters of the
simulated sources helps in validating the results and addressing the issues but does
not enter - directly or indirectly - in the definition of the analysis itself
(see \cite{katz_efficient_2024} for a similar discussion). The ``Sangria''
data contains a simplified noise model (no gaps and no glitches in the data)
with the strongest and most numerous populations of binaries with
relatively weak correlation; it aims at the first exploration of the ``global
fit''. The simulated data contains a population of MBHBs and GBs. The
instrumental noise was simulated as a stationary Gaussian process with the
colour described in \cite{babak_lisa_2021}. Three groups (besides us) have
submitted the results \cite{littenberg_prototype_2023, katz_efficient_2024,
	strub_global_2024} which vary in implementation and sometimes in the
approach (Bayesian in \cite{littenberg_prototype_2023, katz_efficient_2024} vs
frequentist \cite{strub_global_2024}). In this work, we do not make any
comparative analysis, which is an ongoing effort and will be published
separately.

The structure of this paper is as follows. We give a detailed
description of the ``Sangria'' simulated dataset in the next
section. \cref{sec:framework} being the longest section, we start by introducing
the notation and basic notions of data analysis, then
describe how we search for the MBHB and GB signals in
\cref{sec:MBHB_kick_off,sec:GBsearch}. The found sources are fed into an
iterative parameter estimation ``wheel of fortune" where we perform
the block Gibbs-like sampling across different sources of GW and the
noise, described in \cref{sec:gibbs-iterations}. We apply the pipeline
to ``Sangria'' and summarize the results in
\cref{sec:results}. Finally, we discuss the trajectory for the
improvement of the current algorithm and their extension in
\cref{sec:discussion}. Throughout the paper, we use geometrical units \(G=c=1\).
\vfill

\section{Data description}\label{sec:data_desc}
In this short section, we describe the simulated data ``Sangria'' which was
used to validate the data analysis algorithms.

The instrumental noise was generated as a stationary Gaussian process, the
``colour" of the noise was based on the acceleration and optical metrology
noise components (see \cite{babak_lisa_2021} for definition and details),
however, the level for each noise component was chosen within a Gaussian prior
centred on the fiducial values with 20\% of variance. Fitting the level of
instrumental noise is one of the objectives of this challenge.

The GW sources are added one-by-one and we assume the first generation
(TDI-1.5) Michelson measurements (see \cite{babak_lisa_2021}).

The GB population is based on the model given in \cite{korol_populations_2020}
for the detached white dwarf binaries. The interacting binaries are added based
on the estimations in \cite{nelemans_short-period_2004}, which should be
updated in the next LISA simulations. We have also simulated the most
up-to-date set of verification binaries; their live list is maintained in
\url{https://gitlab.in2p3.fr/LISA/lisa-verification-binaries}. At the moment of data
production, we had information about the \(36\) verification binaries:
white dwarf binaries with known sky position and orbital period (the unknown
parameters are simulated from uniform physical priors). In total, the simulated
data contains about 30 mln binaries. Most of those sources are too weak (and
superposed) to be resolved individually, and they form cyclo-stationary stochastic
GW foreground \cite{krolak_optimal_2007} dominating over the instrumental noise
in the range between \SI{0.2}{\milli \hertz} and 5-\SI{6}{\milli \hertz}. The
shape of this foreground is a function of observation time (we resolve more
sources at the high-frequency end of the confusion as more observational data
become available) and of performance of the data analysis algorithm used to
detect and characterise these sources. Characterising this astrophysical
foreground is an essential task of this challenge. The GW signal is represented
by (almost) monochromatic signals, and we assume that all binaries are on
circular orbits. We can infer the frequency evolution for binaries at high
frequencies, otherwise we can only estimate an upper bound. The GW signal is
modulated by the LISA constellation motion in both amplitude (due to the LISA
antenna response function) and phase (Doppler modulation). These modulations
allow us to localise sources in the sky and create (1/year) harmonics around
twice the binary orbital frequency for the Fourier domain representation of the GW
signal (see \cite{krolak_optimal_2007,blaut_mock_2010}).

The population of merging massive black holes is based on the heavy-seed model
with delays (described as ``Q3-d'' in \cite{klein_science_2016}). In the random
draw from multiple realisations of this model, we have got 15 MBHBs. Each
binary was simulated using \texttt{PhenomD} \cite{khan_frequency-domain_2016}
model, which assumes that black hole spins are parallel to the orbital angular
momentum (no orbital precession) and using only a dominant \((2,\pm 2)\) mode
(twice orbital frequency). Note that using only one dominant mode makes it
harder to properly estimate the parameters due to multiple degeneracies in the
parameter space \cite{marsat_exploring_2021}. All these binaries merge during
the one year of the simulated observations and produce GWs with a
signal-to-noise ratio (SNR) ranging between 77 and 3260. Note that \emph{all}
signals could be identified by the eye in the whitened (or band-passed) data. The
estimation of the merger time for each binaries could be easily achieved with a
precision within \num{\sim 500} sec with a few steps of matched filtering. The
accumulation of the SNR for those sources comes from a few hours (for weak and
massive binaries) to a few weeks (for bright and ``less" massive binaries)
before the merger. Therefore, these binaries can be treated as transient GW
signals (as compared to 1 year of observation) and can be analysed in a
few-weeks-long data segments. In the frequency domain, the GW signals from
MBHBs are broadband and (often) stand above the instrumental noise level. We
need to use the data ``in-between" mergers to make a first noise power spectral
density (PSD) evaluation which enters into the likelihood evaluations. We require
just a few days of data to estimate PSD within the analysis frequency band
(\SIrange[range-phrase=-]{0.1}{25}{\milli \hertz}).

\section{Global fit prototype}\label{sec:framework}
The analysis of data can be split into ``search'' for GW signal candidates and
the ``parameter estimation''.

The main objective of the search part is to detect GW signals or identify the
candidates for the weak signals. In this step, we use a maximum likelihood
search for one source at a time.  This step could be seen as an accelerated
burn-in stage for the Bayesian analysis.

The search for MBHBs is described in
\cref{sec:MBHB_kick_off}, and for GBs in \cref{sec:GBsearch}. The found sources
are added to the parameter estimation step, which is based on the Bayesian
approach and is described in \cref{sec:tools}. The evaluation of noise PSD is
performed iteratively together with updates in the characteristics of GW
sources within the global fit pipeline described in
\cref{sec:gibbs-iterations}. Note that the parameterized noise model includes
GW foreground from the unresolved GBs.

\subsection{Data analysis tools}\label{sec:tools}
We assume that the data \(d(t)\) is given by the noise \(n(t)\) and
the superposition of the GW source \(\sum_i s_i(t)\). The data here is presented by the
time-delay inteferometry TDI-1.5 (see \cite{babak_lisa_2021}), moreover, we work
only with two ``noise-orthogonal" combinations usually denoted as \(d=\{A,
E\}\). Note that these combinations have uncorrelated noise with the same
spectral properties (by construction of the simulated data). We assume that the
noise is stationary and Gaussian, which is definitely true for the simulated
instrumental noise. The stochastic foreground due to GBs is cyclo-stationary
\cite{edlund_white_2005}; however, on the MBHB signal's time scale, we can
consider the noise as stationary. The level of GB foreground varies slowly in
time, and, as a result, the spectral shape of the noise in the frequency domain
is a function of the volume of data we analyse. We consider the entire one year
of data for the analysis of the GBs in the frequency domain, so that the
evaluated spectrum is an average over one year, which does not change, and data
again can be considered Gaussian in the frequency domain. As a result, we have
a Gaussian likelihood, and its logarithm is given as
\begin{equation}
	\ln\mathcal{L} = \frac{1}{2} \ln{ \prod_i S_{\text{n}}(f_i)} - \frac{1}{2} \left(d-H | d-H\right),
\end{equation}
where \(S_n\) is the noise PSD at the set of observed frequencies \(f_i = i/T_{\text{obs}}\), with
\(T_{\text{obs}}\) the total observation time, and the matched-filter inner product is defined as
\begin{equation}
	( a | b ) = 4 \Re \int_{f_{\text{min}}}^{f_{\text{max}}}
	\frac{\tilde{a}(f) \tilde{b}^*(f)}{S_{\text{n}}(f)} \dd{f},
\end{equation}
where tilde denotes Fourier transformed data. The combination of deterministic
(and resolvable) sources could be split into GBs and MBHBs:
\(H(t, \vec{\Theta}) = \sum_k H_{\text{GB}}(t, \vec{\Theta}_k^{\text{GB}}) +
\sum_{k'}H_{\text{MBHB}}(t; \vec{\Theta}_{k'}^{\text{MBHB}})\).

We do not know \textit{a priori} the number of GW sources in the data, so the idea is
to perform the search with a subsequent model selection step where we consider
several competing models, say models \(M_n\) and \(M_{n+1}\) which are assumed
that there are \(n\) and \(n+1\) sources in the data correspondingly. The
probability of each model is given (according to Bayesian theorem) as
\begin{equation}
	p(M_i|d) = \frac{p(d|M_i)\pi(M_i)}{p(d)},
\end{equation}
where we always denote the prior probability as \(\pi(.)\) and \(p(d)\) is probability of the
data which plays the role of unknown normalisation, which forces us to consider the Bayes factor:
\begin{equation}
	\label{eq:BF_pair}
	\mathcal{B}_{M_n/M_{n+1}} = \frac{p(d|M_n)}{p(d|M_{n+1})}.
\end{equation}
The Bayes factor is equal to the odds ratio for the uniform prior over the models (which we always assume here).
The evidence \(p(d|M_i)\) for each model is given as
\begin{equation}
	p(d|M_i) = \int p(\vec{\theta} |d, M_i) \pi(\vec{\theta} | M_i) \,d\vec{\theta}.
\end{equation}
The posterior distribution of parameters for each model (parameters of GBs,
MBHBs and the noise model) is evaluated using Markov Chain Monte-Carlo techniques with parallel tempering
(\texttt{m3c2} \cite{falxa_adaptive_2023},
\texttt{lisabeta} (available at {\url{https://gitlab.in2p3.fr/marsat/lisabeta}})) :
\begin{equation}
	p(\vec{\theta} |d, M_i) = \frac{p(d|\vec{\theta}, M_i) \pi(\vec{\theta} | M_i) }
	{p(d|M_i)}.
\end{equation}
Before we proceed we should introduce a general \(\mathcal{F}\)-statistic which is the result of maximization
of the log-likelihood over the coefficients \(a_j\) of the linear model of the signal
\(h(\vec{\theta}) = \sum_j a_j h_j(t, \vec{\theta}')\) (see
\cite{jaranowski_data_1998,prix_targeted_2009,babak_f_statistic_2025})
which can be written as
\begin{equation}
	\mathcal{F} = \frac{1}{2} X_i M^{-1}_{ij} X_j,
\end{equation}
where \(X_i = (d|h_i)\), \(M_{ij} = (h_i|h_j)\).
During the search, we use likelihood (ratio) as a detection statistic.
The introduced above \(\mathcal{F}\)-statistic is the result of analytic maximization
of the likelihood over some parameters, reducing the parameter space's dimensionality.
We will be using \(\mathcal{F}\)-statistic during the search step for both populations of binaries
and will give a detailed description of the GW signals of each type.

\subsection{Search for MBHBs}\label{sec:MBHB_kick_off}
The search for merging MBHBs is relatively trivial for the astrophysical models
used in the simulation of Sangria. Mergers can be seen by the eye in whitened
data, so this is not strictly speaking a detection problem. However, we still
need to quickly assess the merger time and intrinsic parameters of the binary.
This can be done in a ``kick-off'' step where we optimise the model's
parameters (maximum likelihood) and subtract the signal. The MBHBs are the
loudest signals in the data, their broadband nature corrupts the estimation of
noise PSD and hinders the detection of the quasi-monochromatic GB sources.

The suppression of MBHB mergers does not require to be perfect, only to be good
enough to allow the detection of the GB and to start an iterative parameter
estimation process. We, therefore, favour efficiency over accuracy.

To achieve the efficient suppression of MBHBs, we exploit their transient
nature: most of the SNR for these signals comes within a few weeks prior to the
merger. We analyse the data in 2-week long segments. We also make a crude
approximation to the model of the GW signal; in particular, we assume a static
LISA in the response functions and further partially decompose the response
assuming \(\omega L \ll 1\), where \(L\) is LISA armlength and \(\omega\) is GW
angular frequency. Note that, this is in many respects a phenomenological
template waveform, where we care mainly about the accuracy of its phase
evolution, which depends on the intrinsic parameters of the binary: masses and
spins. We are using the \texttt{PhenomD} model which gives GW strain decomposed
in the spherical harmonics. Following
\cite{cornish_lisa_2003,marsat_exploring_2021}, the single link response kernel
of LISA is given as\footnote{Note that there is a typo
	of \(\frac{1}{2}\) in the corresponding equation presented in \cite{marsat_exploring_2021}.}
\begin{equation}
	G^{\ell m}_\text{slr} \approx i \pi fL
	\exp\left(2i \pi f \bf{k} \cdot \bf{p}_0\right) [ {\bf n}_{\text{l}}
	\otimes {\bf n}_{\text{l}}] : {\bf P}_{\ell m},
	\label{eq:G_slr}
\end{equation}
with \({\ell m}\) indices of a particular GW harmonic. We use a multi-harmonic expression
to emphasise the generality of our approach, even though we use only \((2,\pm2)\) harmonics
in the analysis. In addition, \(\bf{k}\) is the the direction of GW propagation, \(\bf{p}_0\)
is the position of the centre of the LISA constellation, \(\bf n_\text{l}\) is the link unit
vector pointing from the sending spacecraft to the receiving spacecraft,
and \(P_{\ell m}\) is the polarization matrix corresponding to \({\ell m}\) harmonic.
With this approximation, the A and E data templates (waveforms) can be expressed in frequency
domain as \cite{marsat_exploring_2021},
\begin{equation}
	\begin{aligned}
		\tilde{A}^{\ell m}, \tilde{E}^{\ell m} :=
		 & i \sqrt{2} \mathrm{sin}(2 \pi f L) e^{-2 i\pi f L} (-6i\pi fL) \\
		 & \times e^{2i\pi f \bf{k} \cdot \bf{p}_0} F_{a,e}^{\ell m}
		\tilde{h}^{\ell m},
	\end{aligned}\label{eq:tdi-mode}
\end{equation}
where \(F_{a,e}^{\ell m}\) are the antenna beam functions respectively for A and E channels,
and \(\tilde{h}^{\ell m}\) is the waveform harmonics determined by the intrinsic parameters,
i.e.\ the masses and the spins of the binary. Two additional parameters are the luminosity
distance defining the amplitude and the merger time. Since we neglect the LISA constellation
motion, the antenna beam functions are just constant terms. Note that we do not expand the
scaling factors in the long-wavelength approximation in
\cref{eq:tdi-mode}.

Strictly speaking, the merger cannot be considered a ``long wavelength
approximation", however we have noticed that modifying \(e^{-2 i\pi f L}\) to
\(e^{-4 i\pi f L}\) improves the agreement of our crude approximation with the
signal that includes a full response and the merger, therefore we adopt this
change in the search.
The resulting template is simple and fast to generate. Note that this signal
also had a desired form for applying the \(\mathcal{F}\)-statistic maximising
over the amplitudes and the constant phase in \(A\) and \(E\) channels. We also
maximise over the merger time using the Fourier transform, the procedure often
used in the ground-based data analysis by the LIGO-Virgo-Kagra collaboration.

The described above procedure allows us efficiently maximize the likelihood over all extrinsic parameters:
distance, sky position, time of coalescence, polarization, initial phase. We reduced the dimensionality of
the search space to four intrinsic parameters (masses and spins).  Using F-statistic (with the time-shift)
we analytically maximize the likelihood over the extrinsic parameters. In this way, we reduce the dimensionality
of the search space to the 4 intrinsic parameters. We used Powell's conjugate direction method \cite{powell_efficient_1964}
to maximize the likelihood function for the run presented in this paper.
In the end, we have the maximum likelihood reconstruction of each MBHB signal within the
limits of the faithfulness of our approximate model. We have also considered an alternative
approach where we run MCMC on the \(\mathcal{F}\)-statistic (as an optimizer)  using an original
long-wavelength approximation and truncating the data in the frequency domain outside of its
validity. Both approaches give comparable results in precision and efficiency.

As a next step, we fix intrinsic parameters at their maximum likelihood
values and run MCMC using a full likelihood and a long-wavelength approximation
to recover the extrinsic parameters. The complete set of parameters is used to construct
the reference waveform in the heterodyning procedure (described in \cref{sec:mbbh_block})
in the first loop of the global iteration procedure.

\subsection{Search for GBs}\label{sec:GBsearch}
The bandwidth of each
GB is very narrow (typically \SI{2}{\micro \hertz}) and we search/analyse these
sources in the frequency domain, splitting the full range in the overlapped
sub-bands, forming a total of 4663 bands.

The search part for GB is closely integrated with the parameter estimation
part, which we describe in \cref{sec:gb_block}. The GW signal from each GB is
characterised by 8 parameters: the central frequency \(f\), the frequency
derivative \(\dot{f}\), the ecliptic latitude \(\beta\), the ecliptic longitude
\(\lambda\), the amplitude \(A\), the inclination angle \(\iota\), the
polarization angle \(\psi\), and the initial orbital phase \(\Phi\). Note that
the signal is decomposable in \(h = \sum_j a_{j} \tilde{h}_j(f,
\theta'_{\rm{GB}})\) where \(j = \overline{1,4}\) and the coefficients \(a_j\) are
functions of the extrinsic parameters (\(A\), \(\iota\), \(\psi\), \(\Phi\)).
The corresponding \(\mathcal{F}\)-statistic depends only on the sky position
(\(\lambda\) and \(\beta\)), the central frequency \(f\) and the derivative
\(\dot{f}\) \cite{blaut_mock_2010}.

We perform the search (maximum \(\mathcal{F}\)-statistic) on a constant grid
similar to the algorithm described in
\cite{blaut_mock_2010,timpano_characterizing_2006}. The grid is generated in a
box delimited by the prior distribution of the intrinsic parameters, and the
found candidate is processed in the parameter estimation part, both described
in detail there. The search is performed in each iteration cycle.
The grid is quite coarse, so (in the spirit similar to \cite{blaut_mock_2010})
the grid search is refined with a short MCMC ran on the \(\mathcal{F}\)-statistic
and by Powell optimisation \cite{powell_efficient_1964}.
The template grid above \SI{18}{\milli \hertz} is quite large, while we expect only a
few sources, so we employ a stochastic search method: adaptive particle swarm
optimisation, APSO \cite{APSO}.

\subsection{Gibbs-like iterations}\label{sec:gibbs-iterations}
The simulated data is analysed in the narrow frequency bands for GBs and in
short time-series segments for MBHBs. We need to glue these two approaches
together in addition to the broadband estimation of the overall noise
(including the GW foreground).

We use a previously suggested approach [\textit{Cornish \& Cutler, private
			communication}, (see also \cite{Cornish_Crowder2005}
		for a discussion of global fit)] performing block-Gibbs sampling across
GW signals of different types
with some important specificity described below. In block-Gibbs sampling, we
assume that the sources of different kinds weakly correlate, so we sample one
type of source at a given step, passing on the residuals to sample the next
type of GW sources in turn. Note that the assumption of the ``weak
correlation'' impacts only the method's efficiency but not its applicability.

In practice, all different source blocks run concurrently, but the noise
block needs to run sequentially.

For each MBHB found in the kick-off step, we initialise a module block and feed
it with the maximum likelihood estimation of parameters. In the next iteration loop,
the reference parameters of each MBHB are set to the last point of the
Markov chain from the previous loop. In addition, we supply
the data as residuals, where all other identified sources are subtracted.
Lastly, we also provide the latest estimate of the noise model used in the
likelihood evaluation. We perform MCMC sampling with parallel tempering and then
record the resulting chain on the data storage.

For each narrow frequency band, we instantiate the GB block module. Similarly,
the input is an initial set of parameters for each source currently considered
in this sub-band, residual (other GWs subtracted) data, and the current noise
model. The search and parameter estimation steps are performed in
this block in turn. The newly identified during the search GB candidates are
added to the model and used in Bayesian parameter estimation. At the exit of
the block, we save the chains on the data storage.

The data, together with the set of parameters at the end of MBHB and GB block
runs, are passed to the noise estimation block, which estimates the parameters
of the noise model.

\begin{figure}
	\centering
	\includegraphics[width=\linewidth]{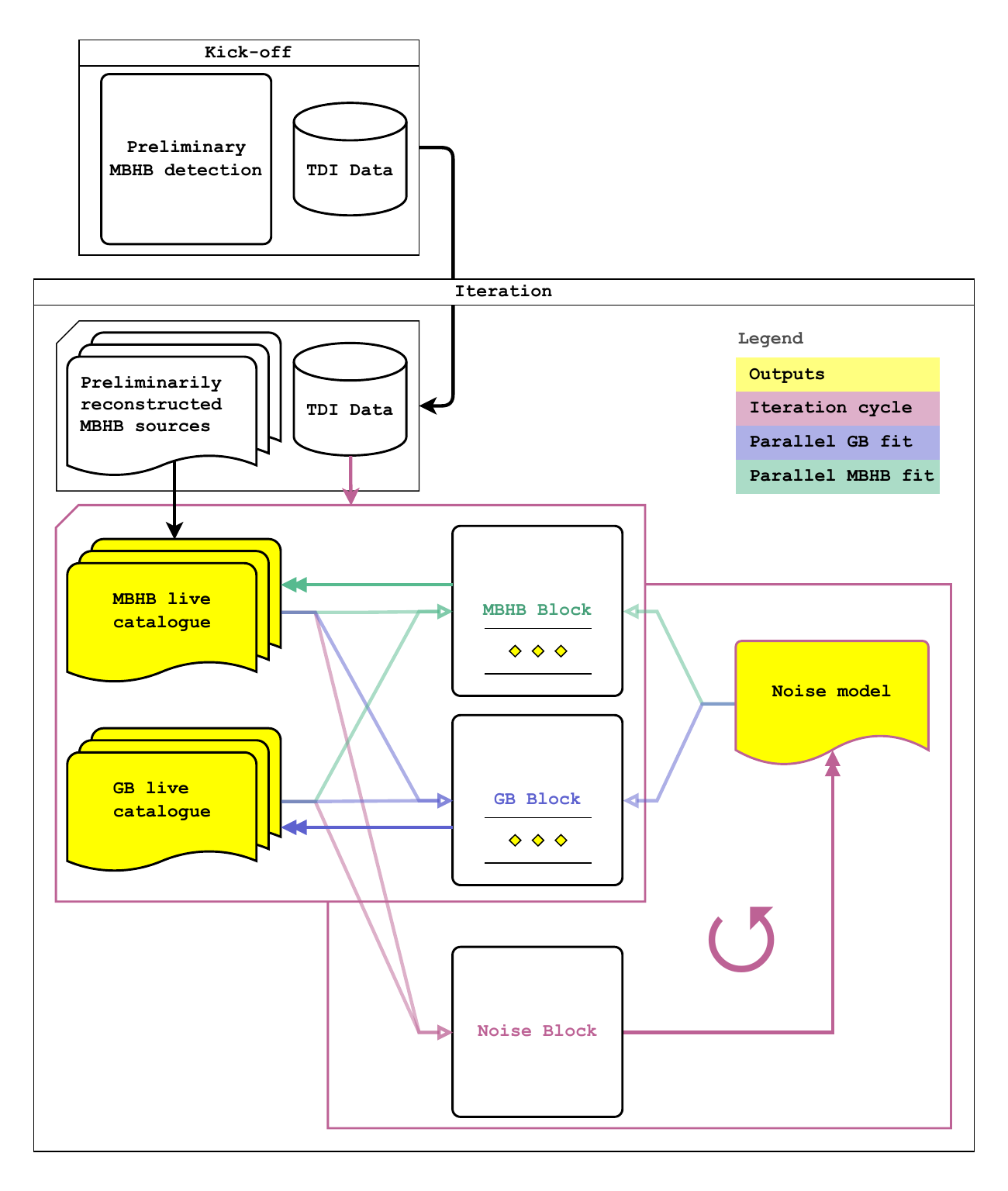}
	\caption{Architecture of the pipeline. Arrows indicate the flow of information.
		Black arrows mean the piece of information is passed only once in the full process,
		whereas the other colours indicate the information is transfered for each iteration cycle.
		Inside each cycle, the MBHB and GB blocks  run concurrently and the noise block  runs
		sequentially after other blocks are finished.
	} \label{fig:globalfit-architecture}
\end{figure}

All chains are saved on the data storage, and the end points are passed as
input to the next iteration. Schematically, it can be presented as
\cref{alg:framework} and \cref{fig:globalfit-architecture}.

\begin{algorithm}
	\KwData{Simulated data}
	\KwHResult{Joint posterior distribution of
		all found sources and noise parameters}
	\BlankLine
	\Begin(MBHB search: kick-off step){
		Search for MBHBs in 2-weeks-long segments;
		\PForEach {data segment}{
			Estimate intrinsic parameters and the merger time
		}
	}
	\Begin(the iterative step){
		Spawn instances of GW source blocks and the noise block\;
		\While{convergence is not achieved}{
			Perform the noise block algorithm\;
			\PForEach{source block instance}{
				Perform the corresponding block algorithm\;
			}
		}
	}
	\caption{Global-fit framework}\label{alg:framework}
\end{algorithm}

Next, we describe each block algorithm in detail.

\subsubsection{Massive black hole binary block}\label{sec:mbbh_block}
The main idea of this block is to perform parameter estimation of a
single MBHB given the data, noise model and initial parameters. In
particular, we use \texttt{lisabeta} and the multi-particle MCMC with
parallel tempering provided therein. In order to reduce the degeneracy in
the parameter space, we transform the extrinsic parameters of the
binary as described in \cite{marsat_exploring_2021}. As extrinsic
sampling parameters, we use the luminosity distance \(D_L\), the
inclination angle \(\iota\), the phase at coalescence \(\varphi\), the
polarization angle \(\psi_\text{L}\), the ecliptic latitude
\(\beta_\text{L}\), and the ecliptic longitude \(\lambda_\text{L}\),
where the subscript L stands for the LISA frame, this is a
non-inertial frame that rotates with the LISA constellation.

We use uniform priors on the chirp mass (\SIrange{e5}{e7}{\solarmass}), mass
ratio (\numrange{1}{5}), luminosity distance (\SIrange{e3}{2e5}{\mega
	\parsec}), individual spins (\numrange{-1}{1}), the priors for phase-angles
parameters are uniform in \([0, 2\pi]\) and polar angles are uniform on the
unit sphere (cosine-uniform).

Each block instance inherits a short data segment from the kick-off step. The
initial point is used for the heterodyning of the signal (\(\hat{h}_\mathrm{ref}\))
\cite{zackay_relative_2018, cornish_heterodyned_2021,cornish_fast_2013} to
improve the efficiency of the sampling. The main idea is to remove an
approximation of the signal that leaves behind only low-frequency residuals that
can be sampled sparsely in the time domain or require a few points in
frequency.

The heterodyned likelihood computation is a two-step process. First, we
introduce a normalized reference waveform template \(\hat{h}_\mathrm{ref}\) in
the frequency domain, that is, \(\hat{h}_\mathrm{ref} = \exp(i
\arg{h_\mathrm{ref}})\), therefore \(\hat{h}_\mathrm{ref}\) carries the same
phase as \(h_\mathrm{ref}\) but does not have zero crossings. A
template waveform is usually generated on a sparse frequency grid and then
interpolated to a dense grid. The interpolation algorithm consists of the
choice of a family of basis functions and the computation of the interpolation
coefficients corresponding to the basis functions. The support of each basis
function contains a few sparse frequency grid points, and the union of the
supports of all basis functions covers the entire frequency range. Importantly,
for a given interpolation algorithm, the basis functions are entirely
determined by the set of sparse frequency grid points. The latter is fixed once
the reference waveform template is chosen. As a preparation, we compute the
matched filter product between the data and \(\hat{h}_\mathrm{ref}\) for each
basis function with the dense grid covered by its support. We also evaluate the
product of \(\hat{h}_\mathrm{ref}\) with itself for each basis function. This
step is performed only once for each reference waveform template. Second, to
compute the likelihood, we evaluate the template waveform \(h\) on the
sparse reference grid. Then we collect the interpolation coefficients for the
ratio \(h / \hat{h}_\mathrm{ref}\) and the result is given by a combination of
these interpolation coefficients and the precomputed products. This procedure
prevents costly waveform generation on the dense frequency grid.

\subsubsection{Galactic binary block}\label{sec:gb_block}
The GBs are the most numerous sources in the simulated data. This block type
aims at resolving as many GBs as possible, leaving the rest of the population
as a confusion noise contributing to the overall noise budget.

Each block focuses on the analysis of GB in narrow frequency bands with
width \SIrange[range-phrase=-]{1}{10}{\micro \hertz} within the overall band
from \SIrange{0.1}{21.5}{\milli \hertz}. The width of each band is a function
of the central frequency, chosen according to the expected signal width and the
expected density of the GB sources based on a model of the Galaxy, and we aim
at having fewer than 8 sources per band. We also take into account the
bandwidth of the GB signals, which is larger at the high end of the LISA's
sensitivity due to a measurable frequency drift. Note that our algorithm does
not depend on this choice and bands could be subsplit if needed between
iterations. Limiting the number of sources per band reduces the dimensionality
of the parameter space and improves the convergence rate.

At each iteration, we first search for a new GB candidate in the residual data,
i.e.\ the in-band data with already identified MBHBs and GBs in the central
live catalogue subtracted. After the search, ideally, we would employ a
Bayesian model selection scheme to simultaneously estimate the number of GB
sources (taking into account the identified candidates) and perform the parameter
estimation. However, we have realised that the model selection algorithm
severely suffers from the lack of convergence if we have models containing more
than 3-4 GBs. This is a technical issue related to the efficiency of the PTMCMC
sampling in the multidimensional parameter space. The current computational
cost is very high, and we decided to complete a full Bayesian exploration after
improving efficiency (using, for example, normalising flow \cite{korsakova_neural_2024}
and reweighting techniques).

In this work, we employ the hybrid approach: we use a frequentist approach to
determine the GB candidates (the search part), and then we perform the Bayesian
parameter estimation for the identified sources. In the improved version, we will
use the frequentist step described below as a burn-in stage for the fully Bayesian approach.
In the Bayesian approach, we will use the product-space technique \cite{hee_bayesian_2015,carlin_bayesian_1995} to evaluate
the Bayes factor \cref{eq:BF_pair} for pair-wise models. This improvement is already implemented,
but suffers from the convergence issues mentioned above; therefore, we do not present here the partial results.

Within the frequentist approach, we follow the Neyman–Pearson viewpoint \cite{lrrKrolak}, where
we compute the detection threshold based on the desired false alarm
probability. We use \(\mathcal{F}\)-statistic (which is the log-likelihood
ratio maximised over extrinsic parameters) as a detection statistic, which follows
the \(\chi^2\) distribution with 4 degrees of freedom for a single GB
\cite{blaut_mock_2010} in the absence of a signal.
Following \cite{lrrKrolak}, we identify the number of independent
trials required to cover the parameter space as \(N_\text{cell} = V /
V_\text{cell}\) where \(V\) is the volume of the parameter space and
\(V_\text{cell}\) is the volume occupied by a signal and is defined through the
Fisher information matrix as
\begin{equation}
	\label{Eq:sig_vol}
	\Gamma_{ij} \delta \theta^i \delta\theta^j = 1/2,
\end{equation}
where \(\Gamma_{ij}\) is the reduced
(4-dimensional for a single GB) matrix. The components of the reduced Fisher matrix are approximately constant \cite{blaut_mock_2010}:

\begin{equation}
	\Gamma =
	\begin{pmatrix}
		\displaystyle \frac{\pi^2}{3} & \displaystyle \frac{\pi^2}{6}    & 0                                & \displaystyle -\frac{1}{n}  \\[1em]
		\displaystyle \frac{\pi^2}{6} & \displaystyle \frac{4\pi^2}{45}  & \displaystyle \frac{1}{2\pi n^2} & \displaystyle -\frac{1}{2n} \\[1em]
		0                             & \displaystyle \frac{1}{2\pi n^2} & \displaystyle \frac{1}{2}        & 0                           \\[1em]
		\displaystyle -\frac{1}{n}    & \displaystyle -\frac{1}{2n}      & 0                                & \displaystyle \frac{1}{2}
	\end{pmatrix},
\end{equation}
where \(n\) is the observation time in years (\(n=1\) in our case). We have used parametrisation
\(\theta^i = \{f, \dot{f}, A, B \}\) where
\begin{equation}
	\begin{aligned}
		A & = 2\pi f R \cos{\beta} \cos{\lambda}, \\
		B & = 2\pi f R \cos{\beta} \sin{\lambda},
	\end{aligned}
\end{equation}
and \(R\) is 1 AU. In this case, the \cref{Eq:sig_vol} becomes the equation of a 4-dimensional ellipse and
we can evaluate its volume as
\begin{equation}
	V_\text{cell} = \frac{(\frac{\pi}{2})^2}{2\sqrt{\det \Gamma}},
\end{equation}
(see \cite{lrrKrolak} for higher dimensional case).
The total volume is defined by the limits of the parameter space: \(f\) varies within \([f_\text{min}, f_\text{max}]\),
\(\dot{f}\) in \([\dot{f}_\text{min}, \dot{f}_\text{max}]\), \(\beta\) in
\([-\pi/2, \pi/2]\) and \(\lambda\) in \([-\pi, \pi]\) and is given by
\begin{equation}
	V = \frac{2}{3} \pi R^2 \left( f_\text{max}^3 - f_\text{min}^3 \right)
	\left( \dot{f}_\text{max} - \dot{f}_\text{min} \right).
\end{equation}
It gives us  the number of independent trials: \(N_\text{ind} = \min(N_\text{cell},
N_\text{trials})\), where \(N_\text{trials}\)  is an actual number of trials used during the search.
The false alarm probability is then given by
\begin{equation}
	P_\text{FA}(\mathcal{F}) = 1 - \mathrm{CDF}(\mathcal{F})^{N_\text{ind}},
\end{equation}
where CDF is the cumulative distribution function of the \(\chi^2\) distribution
with 4 degrees of freedom. We set the threshold for the false alarm probability
to 1\%, which gave us the detection threshold. Only candidates
above this threshold are retained for the Bayesian parameter estimation.
The exception is the frequency	bands near the knee frequency
\(f_\text{knee} \approx 3\)-\SI{5}{\milli \hertz} (see \cref{sec:noise_block})
where the data is highly non-Gaussian. This band corresponds to the combined GW
signal produced by the GB population losing its stochasticity and the Bayesian model selection
is really necessary. In this frequency range, we postulate the SNR threshold \(\rm{SNR} \ge 6\)
for the detection and keep sources with \(\rm{SNR} \ge 4\) as possible candidates.
Let us reiterate again that the
future version of the pipeline will treat the number of sources as a
random variable and the found candidate will set the dimensionality of the
considered models and a starting point.

In the parameter estimation we use a narrow frequency prior around the
frequency of a found candidate (\(f_*\)): \(f\) is uniformly distributed in
\([f_* - \Delta f, f_* + \Delta f]\) where \(\Delta f\) is the width of three
Fourier frequency bins. We deliberately narrow down the prior range for \(f\)
to alleviate the label switching problem (another way to address it is
described in \cite{buscicchio_label_2019}) and to improve the efficiency of the
MCMC sampling. As long as the posterior distribution does not touch the
boundaries, this narrow prior does not affect the inferred posterior
distribution. The prior distribution for the frequency derivative \(\dot{f}\)
is uniform in \([\dot{f}_\text{min}, \dot{f}_\text{max}]\), where
\(\dot{f}_\text{min}\) and \(\dot{f}_\text{max}\) are respectively given by
\(\num{e-5} f_\text{min}^{5}\) and \(\num{2.25e-6} f_\text{max}^{11 / 3}\) with
\(f_\text{min/max}\) the boundaries of the block frequency band in \si{\hertz}.
The ecliptic latitude \(\beta\) and longitude \(\lambda\) are uniformly
distributed on the sphere. The amplitude \(A\) is logarithmically distributed
in \([10^{-24}, 10^{-20}]\). The inclination angle \(\iota\) is cosine-uniform
in \([0, \pi]\), the polarization angle \(\psi\) is uniform in \([0, \pi]\),
and the initial orbital phase \(\Phi\) is uniform in \([0, 2\pi]\). We use our
MCMC sampler (\texttt{m3c2}) with parallel tempering for parameter estimation.
The identified GB candidates  are used for intialising the MCMC sampler.
Once we consider the chain has converged, we update the
live catalogue and save the chain to the data storage.

\begin{figure}
	\centering
	\includegraphics[width=\linewidth]{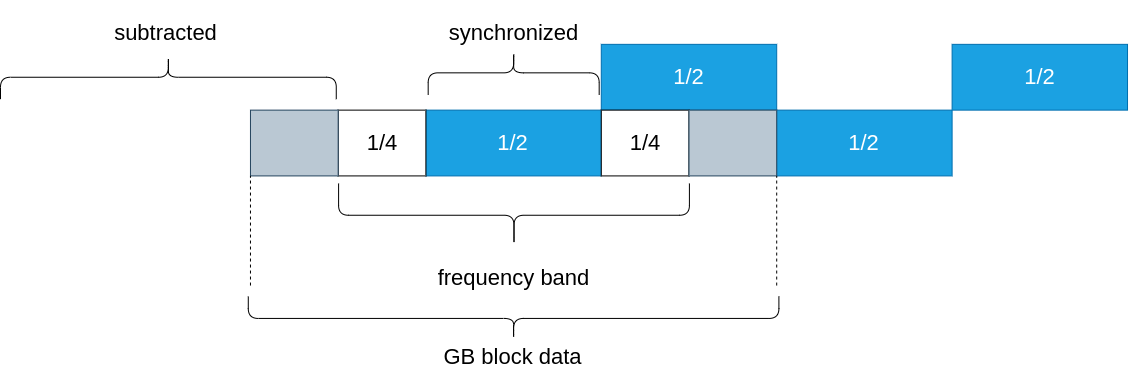}
	\caption{The narrow frequency band in the GB block. The shown data contains 3
		GB blocks (``GB block data''), each composed of a core (blue tile) and two wings
		(white and grey tiles). The core takes half of the width of the band in the centre.
		During the data preparation of a GB block, GB signals in the central live
		catalogue that belong to the neighbouring but not the current and the adjacent
		core bands are subtracted. Analyses are performed in the full band (GB block data)
		but only the results with the central frequency in the core are synchronised to
		the central live catalogue and the central storage of the chains.}
	\label{fig:gb-bands-overlaps}
\end{figure}

We need to give special consideration to sources that fall within the boundaries that separate the
frequency bands. We split the whole data frequency range into a set of
frequency band ``cores" (blue tiles in \cref{fig:gb-bands-overlaps}.). To each
core, we add two ``wings" extending to the center of the adjacent cores (white
and grey tiles in \cref{fig:gb-bands-overlaps}) to take into account the
frequency spread (width) of the GB signals. To prepare the data for the given
GB block, we subtract all identified MBHBs and GBs from neighbouring bands
using the source parameters stored in the live catalogue. In this way, we prevent
contamination of the current block by the neighbouring signal leaking into it
and avoid double-counting of the same sources.
At the end of the parameter estimation step, we only synchronise the
GBs of central frequency in the core (blue tile) to the central live catalogue.
The GBs in the wings (white and grey tiles) are only saved for housekeeping
so that we can understand what we see in the whole GB block. A source falling on the boundary
between blocks might be detected twice in
adjacent blue tiles. To avoid double counting, we systematically check the
presence and uniqueness of found signals before synchronising to the central
live catalogue by computing the overlap between GBs found in the neighbouring
bands. We considered two sources to be the same if their overlap is greater than 0.99.

\begin{algorithm}[t]
	\KwData{L1 data \(D_\mathrm{L1}\)}
	\KwHResult{Joint posterior distribution of
		the GB sources present in the frequency band}
	\BlankLine
	\(H_\text{MBHB}\): Waveform of the MBHBs in the live catalogue\;
	\(H_\text{GB, adj}\): Waveform of the GBs in adjacent bands in the live catalogue\;
	Get \(D \leftarrow D_\mathrm{L1} - H_\text{MBHB} - H_\text{GB, adj} \)
	and restrict to the frequency band\;

	\For{iteration \(i\)}{
		\(k\): Number of GB candidates from iteration \(i-1\)\;
		\(H_\text{GB, i-1}\): Waveform of the \(k\) candidates\;
		\tcc{New source search}
		\Begin{
			\KwData{Residual data \(R \leftarrow D - H_\text{GB, i-1}\)}
			Maximize \(\mathcal{F}\)-statistic in the parameter space\;
			\If{maximum above threshold}{
				Number of GB candidates: \(k \leftarrow k+1\)\;
			}
		}
		\tcc{Parameter estimation}
		\Begin{
			\KwData{\(D\)}
			Run \texttt{PTMCMC} for \(k\) GBs template\;
			Save the chain\;
			Update the live catalogue\;
		}
	}
	\caption{GB block algorithm.}\label{alg:algogb}
\end{algorithm}

\subsubsection{Noise model block}\label{sec:noise_block}
The noise level can be decomposed into two parts: instrumental noise and confusion
noise made of the superposition of unresolved GB sources, as we
mentioned in \cref{sec:data_desc}. The GW foreground from GBs dominates over the
instrumental noise between 0.2 and 5-\SI{6}{\milli \hertz}. The exact shape of the noise in this range is
a function of the algorithm used to identify the resolvable GBs (and MBHBs) and
also of the data volume: the SNR of the GBs grows approximately as
\(\sqrt{T_{\rm{obs}}}\), where \(T_{\rm{obs}}\) is the observation time, so we resolve
more sources as we accumulate the data.

In our analysis, we directly deal with one year of observations, so we are
modelling the average (over the year) noise PSD. We have abandoned the idea of
fitting the confusion foreground independently in the narrow bands because it
is susceptive to very large fluctuations, instead, we impose a smooth
broad-band noise model. The model has 7 parameters, 2 accounting for the
instrumental noise and 5 modelling the confusion noise. Namely, the
instrumental noise parameters are acceleration noise \(S_{\mathrm{acc}}\)
(\cref{eq:test-mass-psd}) and optical metrology noise \(S_{\mathrm{oms}}\)
(\cref{eq:optical-path-psd}). The GB confusion noise parameters
(\cref{eq:gal-noise}) are the amplitude \(A\) defining the overall scale and
directly related to the population model of the GB, the knee frequency
\(f_{\text{knee}}\) defines the transition from the resolvable to stochastic
regime, \(f_2\) is a scaling frequency, and \( (f_1, \alpha) \) define deviation
from the power law. Note that \(f^{-7/3}\) is what is expected from any
population of GW-driven binaries on the circular orbits, while the exponential
factor accounts for resolving the sources at high frequencies, and the
hyperbolic-tan cut-off ends the stochasticity of the foreground. The ``default"
LISA as described in the science requirement document (see also
\cite{babak_lisa_2021}) specifies \(f_{l1} := \SI{4e-4}{\hertz}\), \(f_{u1} :=
\SI{2e-3}{\hertz}\), \(f_{l2} := \SI{8e-3}{\hertz}\), where \(x := 2 \pi f L\).
The Galactic foreground \cref{eq:gal-noise} is based on the empirical
expression suggested in \cite{karnesis_characterization_2021}.

\begin{widetext}
	\begin{align}
		S_{\text{pm}}=S_{\text{acc}}\left(1+\left(f_{l1} / f\right)^2\right)\left(1+
		\left(f / f_{l2}\right)^4\right) \left/ (2 c \pi f)^2 \right.,
		\label{eq:test-mass-psd}                                                              & \\
		S_{\text{op}}=S_{\text{oms}}\left(1+\left(f_{\mathrm{u} 1} / f\right)^4 \right)
		(2 \pi f / c)^2, \label{eq:optical-path-psd}                                          & \\
		S_{\text{gal}}= \frac{A}{2} f^{-\frac{7}{3}} \exp
		\left(-\left(\frac{f}{f_1}\right)^\alpha\right)
		\left(1+\tanh \left(\frac{f_{\text{knee}}-f}{f_2}\right)\right). \label{eq:gal-noise} &
	\end{align}
\end{widetext}

\begin{figure*}
	\centering
	\includegraphics[width=\linewidth]{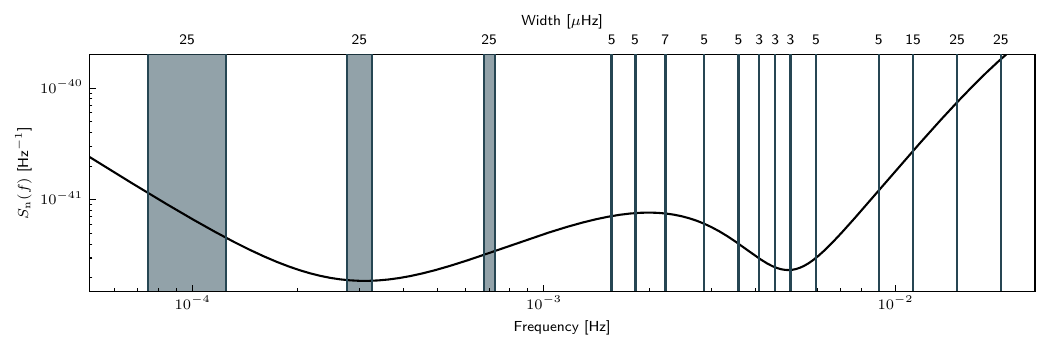}
	\caption{The 16 frequency bands used for the noise model sampling
		with their location and size. The solid black line shows an expectation
		based on \cite{karnesis_characterization_2021}.}\label{fig:noise_bands}
\end{figure*}

These noise components contribute to the PSD of the A and E channels as (see
\cite{prince_lisa_2002} for more details)
\begin{widetext}
	\begin{align}
		S_{\text{instr}}^{A,E} & = 8 \sin^2(x)
		\left(2 S_{\text{pm}}\left(3+2 \cos(x)+\cos(2 x)\right)
		+S_{\text{op}}\left(2+\cos (x)\right)\right), \label{eq:tdi-a-e-instr-noise} \\
		S_{\text{gal}}^{A,E}   & = 6\left(x \sin (x)\right)^2 S_{\text{gal}}.
		\label{eq:tdi-a-e-gal-noise}
	\end{align}
\end{widetext}
The noise block reads the parameters of all GW sources from the previous
iteration and removes them. We sample the noise parameters using parallel
tempering MCMC (\texttt{m3c2}) in 16 narrow frequency bands spread out in the
range \SIrange{0.1}{20}{\milli \hertz} and presented in \cref{fig:noise_bands}.
The bands are chosen so that we have only a few GBs (if at all) in each of
them, and they are sufficient to sample the shape of the expected PSD.
The chosen
number of bands allows us to efficiently obtain a robust result that depends weakly
on a particular choice. The implemented scheme easily allows for a
larger number of bands and different choices of their location. We are
currently also exploring an agnostic model based on piece-wise cubic spline.

Care should be taken to ensure the correctness of the parallel Gibbs sampling
\cite{gonzalez_parallel_2011}. Whereas there is no interdependence of the
source parameters \(\vec{\Theta}^{\text{GB,MBHB}}\) across different time (MBHBs) and frequency (GBs) bands,
the noise model parameters \(\vec{\Theta}^{\text{noise}}\) are coupled to each
other and to all sources. To guarantee the correct convergence, the
sampling of different sources could be performed in parallel, conditional on
the noise model parameters, which need to be updated sequentially.

\subsection{Implementation}
GW sources and noise blocks are encapsulated into an arbitrary number of jobs
and run through a batch system. Jobs are automatically spawned at the end of
each iteration of the Gibbs sampling, and the disk storage is used to exchange
data, with particular care to avoid concurrent access.

To implement global fit, we opted for a coding strategy that favoured
simplicity and flexibility. The entire code is written in Python (although some
third-party dependencies, such as \texttt{lisabeta}, may contain C/C++ code)
and the block architecture has been designed to maximise the degrees of freedom
available for execution. Each block is configured by a \texttt{YAML} file and performs a
specific task (e.g. sampling of a particular type of GW source or the noise)
with an execution routine naturally organised in repeated iterations. One or
more blocks (and iterations of blocks) are executed in batch-type jobs in
series or parallel. Each job is run by the batch system independently on one or
more cores of a single node. The pipeline requires a low rate of information
exchange. All data exchange and task flow control (barriers, dependencies,
etc.) rely on the shared filesystem. Our system is designed to be highly
adaptable. It can be executed on any medium-sized computing centre, whether a
High-Performance Computing (HPC) or High-Throughput Computing (HTC) facility.
Importantly, it does not require any specific hardware, such as an Infiniband
network or GPU, making it accessible to a wide range of users. It also leaves
complete freedom to redefine the workflow at will according to the needs
dictated by the global fit R\&D. \section{Results}\label{sec:results}
In this \namecref{sec:results}, we summarize the results of the analysis of the
``Sangria'' dataset with the prototype implementation of the global-fit pipeline. We ran
the pipeline on the computing cluster provided by CNES (Centre National
d'Études Spatiales), using 1000 cores.

We have performed 10 ``global'' iterations for about 20 hours.
Obviously, it is not enough to get the full
convergence across the noise and all sources. The tails of the posterior distributions are somewhat
truncated. However, we can still assess the results if we are converging to the right solution
and how many sources we managed to identify correctly. We observe that the global burn-in is achieved within
\(\approx 8\) iterations, after that we explore the correlations across the noise and astrophysical populations
and posteriros of individual sources.

\subsection{Merging massive black hole binaries}\label{sec:mbhb_results}
\begin{figure}
	\centering
	\includegraphics[width=\linewidth]{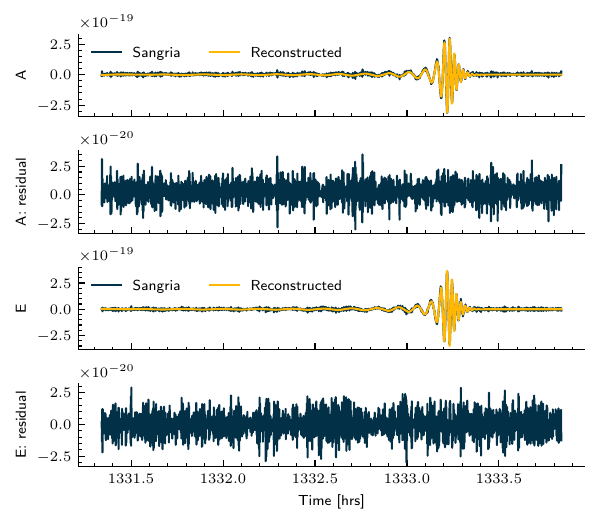}
	\caption{Time series signal of the Sangria dataset zoomed in around the merger
		of the MBHB 0 and the reconstructed signal for that MBHB.
		In the second and the fourth panels (from the top),
		we plot the residuals after the removal of the reconstructed signal
		in TDI A and TDI E channels.
		Note that the data is not whitened in this plot.  }
	\label{fig:kick-off-mbhb0}
\end{figure}

\begin{figure}
	\centering
	\includegraphics[width=\linewidth]{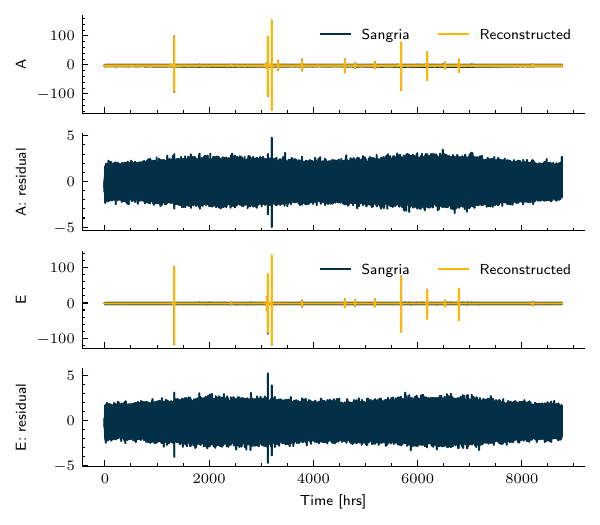}
	\caption{Whitened time series signal of the full Sangria dataset and the
		reconstruction of all the 15 MBHBs. In the second
		and the fourth panels, we plot the whitened residuals after the removal
		of the reconstructed signals in TDI A, E.}\label{fig:kick-off-full}
\end{figure}

The search for MBHB was the first step of the analysis and is described in
\cref{sec:MBHB_kick_off}. The results of the initial MBHB fit are summarised in
\cref{fig:kick-off-mbhb0,fig:kick-off-full}. \cref{fig:kick-off-mbhb0} shows
the reconstruction and residuals after the subtraction of the very first (since
the beginning of simulated observations) MBHB, the time series (\(A\) and \(E\))
are zoomed around the merger time. \cref{fig:kick-off-full} demonstrates
whitened original time series and residuals after subtracting the best
estimates. Let us emphasise that the main aim at this point is to bring the residuals below the noise level,
allowing for better PSD estimation and the
start of the iterative global fit process. This process takes about 1 hour using 1 core
per MBHB.

The effectiveness and quality of the MBHB search step is confirmed by a quick
convergence in the MBHB parameters and in the noise model (see \cref{fig:noise}). Estimation of
the parameters of the MBHB is performed by iterating the
residual data and the noise model as described in
\cref{sec:gibbs-iterations,sec:mbbh_block}. Each MBHB block run accumulates
between five and ten thousands of MCMC samples. This number warrants the minimum
convergence within a given iteration step and balances the synchronisation across
different blocks.  Regarding the convergence, we have targeted the Gelman-Rubin statistic
\cite{gelman_inference_1992} \(R \leqslant 1.2\)
between all the walkers in the
chain. We do not claim that the chains are converged even when we reach this
condition; it is not required at each iteration, as we need a \emph{global}
convergence, which takes into account variations in the residuals of the subtracted GBs and the
noise uncertainties; in other words, we need many ``global'' interactions.
However, we should still allow the chain to ``adjust'' to the new reference waveform in
heterodyning and to new ``noise'' combined out of the noise and varying residuals from
subtraction of other sources.
Each MBHB block runs in parallel for about half an hour within a single iteration loop.

\begin{figure*}
	\centering
	\includegraphics[width=0.48\linewidth]{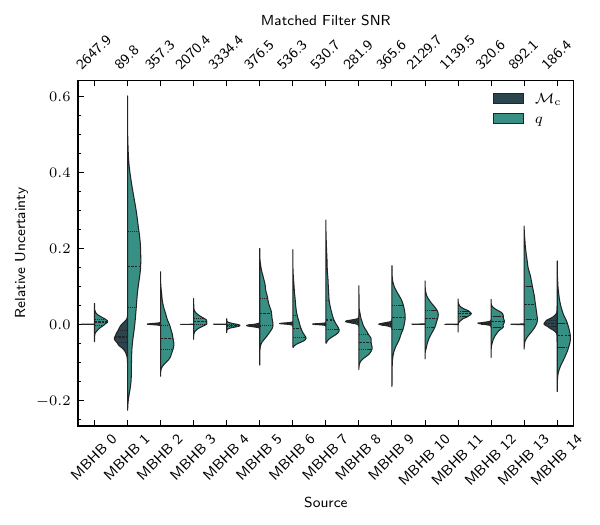}
	\includegraphics[width=0.48\linewidth]{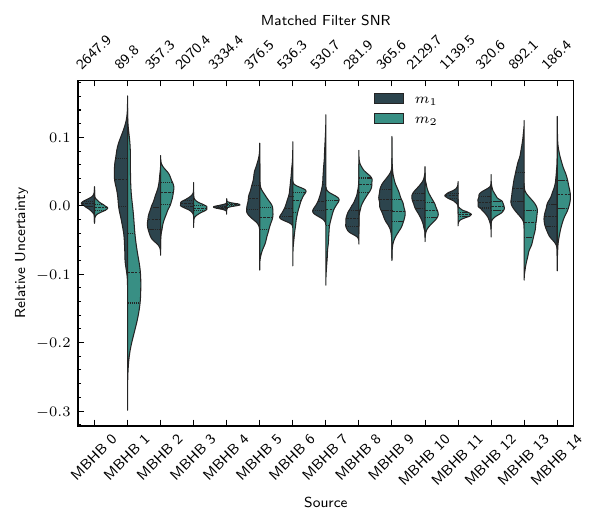}
	\caption{Marginalised posterior distribution of the chirp mass and the mass ratio (\textit{Left});
	individual masses  (\textit{Right})
	for the MBHB sources in the ``Sangria'' dataset. We plot a relative uncertainty
	and zero corresponds to the true parameter values of injected signals. The left plot:
	the left histogram (blue) is for the chirp mass
	\( \mathcal{M}_\text{c} / \mathcal{M}_{\text{c,tr}} -1 \), the right (green) histogram
	corresponds to the
	mass ratio \(q / q_\text{tr} - 1\). The right plot:  \( m_i / m_\text{i,tr} -1 \),
	the blue (left) for the primary (heavier) MBH, the green (right) is for the secondary.
	The dashed lines indicate the quartiles of the distribution.}\label{fig:violin-chirp-ratio}
\end{figure*}

\begin{figure}
	\centering
	\includegraphics[width=\linewidth]{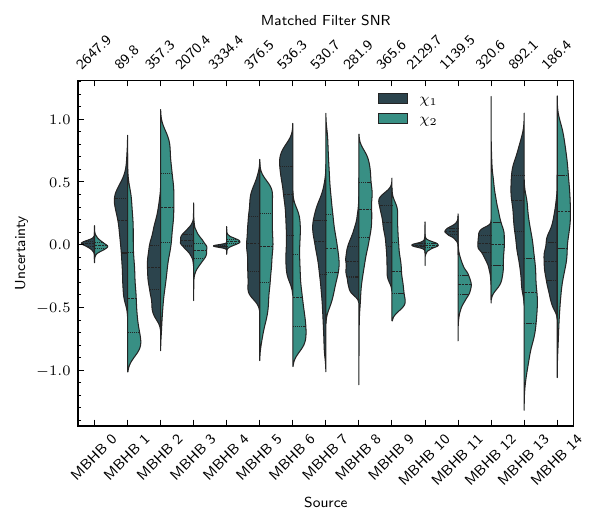}
	\caption{Marginalised posterior distribution of the spin projections for the MBHB sources
		in the Sangria dataset (settings and colours are similar to the right plot of
		\cref{fig:violin-chirp-ratio}). The dashed lines indicate the quartiles of the distribution.
		The \(y\)-axis is the spin projection uncertainty,
		defined as \(\chi_{i} - \chi_{i, \text{inj}}\).}\label{fig:violin-spins}
\end{figure}

To illustrate the performance of the MBHB block, we have done five extra
iterations where we draw GB and noise parameters from the distribution of the last global iteration.
We merge the chains from the last global iteration and the five extra iterations of the MBHB block
to produce the final posterior distribution. The relative uncertainty \((\theta -
\theta_{\rm{tr}})/\theta_{\rm{tr}}\), where \(\theta = \{\mathcal{M}_c, q=m_1/m_2
\}\), \(\mathcal{M}_c\) is the chirp mass and \(\theta_{\rm{tr}}\) is a true value
of the parameter, is shown in the left panel of \cref{fig:violin-chirp-ratio}.
The matched filter SNR for each source \(\sqrt{\left( d | h \right)}\), is
given at the top of the figure. One can see that the chirp mass is the
best-determined parameter, often at the sub-percent level. Astrophysists might
be more interested in the measurement of individual masses which we also plot
in the right panel of \cref{fig:violin-chirp-ratio}. We want to point out a few
interesting features; first, there is a relatively strong bias in the chirp
mass (and individual masses) for MBHB~1, MBHB~8; and second, we observe a
significant bias in the mass ratio for MBHB~2, MBHB~11, MBHB~13. The bias is
very pronounced in the determination of individual masses of MBHB~11. Other
participants in the ``Sangria'' data
challenge \cite{littenberg_prototype_2023} demonstrated similar results, and we attribute it to a
combination of a particular noise realisation and residuals after subtracting
the GBs, but detailed analysis of all submissions is still ongoing. Note that
in all cases (except the weakest MBHB~1), we measure masses within 10\% accuracy.
The marginalised posteriors for individual spins (note that we have used the
model with spins parallel to the orbital momentum) are given in
\cref{fig:violin-spins}. We can again see a significant bias for MBHB~11. The spin
of the primary MBH, as expected, is determined better, usually within 0.5, and
the accuracy is strongly correlated with the SNR of the GW signal.

\subsection{Galactic binaries}\label{results:GBs}
We describe only the detection of GBs in this paper, leaving the results of the Bayesian
inference to the future publication (due to the lack of full convergence).

\begin{figure}
	\centering
	\includegraphics[width=\linewidth]{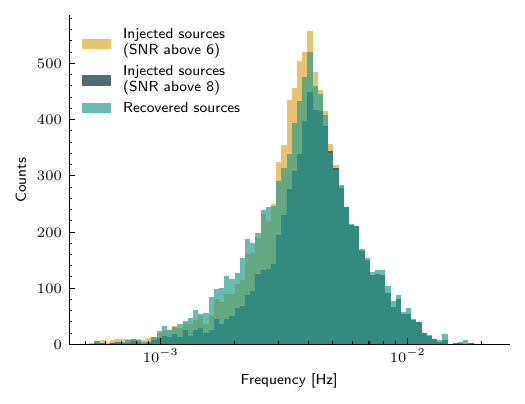}
	\caption{The distribution of detected GBs (cyan) with SNR above 6 compared
		to the catalogue (injected) sources with SNR above 6 (yellow) and 8 (green).
		We use the noise PSD estimated from the global-fit pipeline to compute the SNR. }
	\label{fig:gb-counts-freq-snr-above-8}
\end{figure}

During the analysis, we detected a total of 9542 GBs with SNR above 6. The distribution of
the detected sources across the frequency is shown in \cref{fig:gb-counts-freq-snr-above-8}
in light green; for comparison, we have also plotted the catalogue GBs with SNR above 6
(yellow) and 8 (dark green). This plot shows that we are losing GBs in this SNR range and
in the frequency band where the GW signal from the population of GBs is losing
stochasticity and we need Bayesian inference on the number of sources.
The SNR is computed with the noise PSD estimated from the global fit
pipeline, namely, using the chain's last point from the last iteration. Their
identification with the catalogue sources is not always straightforward; we
have used overlap as a main criterion. We assumed that the GBs are correctly
identified if there is a source in the catalogue with more than 90\% overlap,
and partially recovered if the overlap is between 50\% and 90\%, otherwise we
claim it as a false detection. Note that, in general, it is not a false detection in the sense of the
``noise-generated feature''; in most cases it is incorrect recovery of the true sources in the
simulated data. The two main reasons causing this are (i) getting stuck at the secondary maximum and
(ii) the wrong interpretation of the multiple strongly correlated sources; sometimes,
the former leads to the latter.

\begin{figure*}
	\centering
	\includegraphics[width=\linewidth]{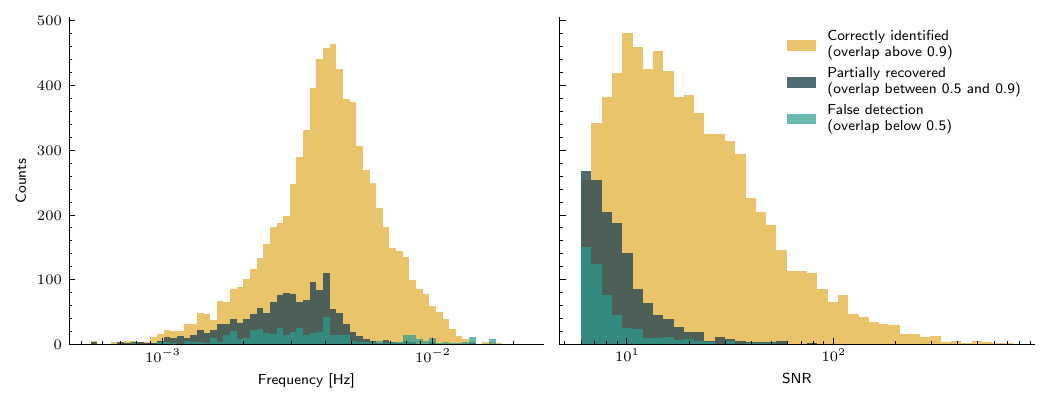}
	\caption{The distribution of identified GBs (yellow),
		partially recovered (green) and false detections (cyan) in
		frequency (left panel) and SNR (right
		panel).}\label{fig:gb-counts-freq-snr}
\end{figure*}

\begin{figure}
	\centering
	\includegraphics[width=\linewidth]{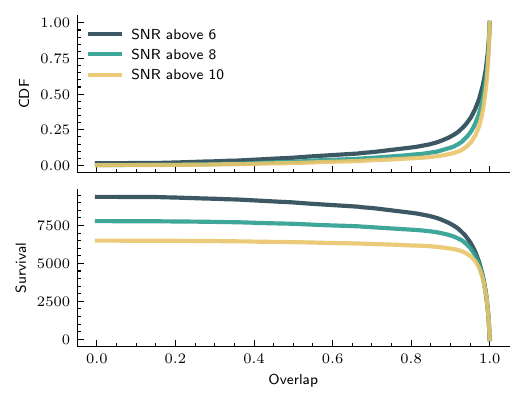}
	\caption{Upper panel:  CDF of detected GBs with three SNR cuts (6, 8, 10)
		as a function of their overlap with the catalogue sources. Lower panel: we plot the
		survived GBs with the same SNR cuts as a function of the overlap.
		A similar plot is given in \cite{katz_efficient_2024}.}\label{fig:gb-pdf}
\end{figure}

Among 9542 detected GBs, we have 7620 identified GBs, 1406 partially recovered GBs, and
516 false detections, their frequency and SNR distribution are presented in
\cref{fig:gb-counts-freq-snr}. Most partial and false detections come from the
source-confused region between 1 and 4 mHz. We also present the number of survived
GBs and the cumulative distribution as a function of overlap in \cref{fig:gb-pdf}.
The main reason for the poor recovery of some bright GBs is the
sampling/convergence of PTMCMC, where we should have used a higher
temperature (and a larger number of parallel chains) in parallel
tempering. In the improved pipeline, we intend to use a combination of
stochastic optimisers such as APSO \cite{APSO} and differential evolution
\cite{DE} to search for isolated GBs at high frequencies.

All in all, we consider the performance of the implemented algorithms to
be very satisfactory. The main aim of this paper and the initial
development is to demonstrate that the implemented method works. We
did not reach full convergence for GBs across the whole frequency
range, but we correctly recovered 85\% of all simulated GBs with
\(\rm{SNR} \ge 8\). We believe that full convergence is unnecessary
at this stage because we continue to work on improving the efficiency of the
sampler and refining the convergence criteria. Having said that, we
have found fully converged posteriors in the frequency bands containing
a few GBs.

\subsection{Noise estimation}\label{results:noise}
\begin{figure}
	\centering
	\includegraphics[width=\linewidth]{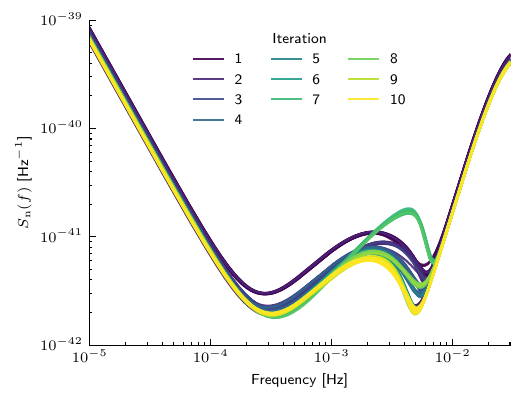}
	\caption{Evolution of the noise power spectral density (PSD) for 10 first
		iterations of the global fit. For each iteration, we plot a dozen curves
		corresponding to the randomly chosen sample points (noise model parameters).
	}\label{fig:noise}
\end{figure}

\begin{figure}
	\centering
	\includegraphics[width=\linewidth]{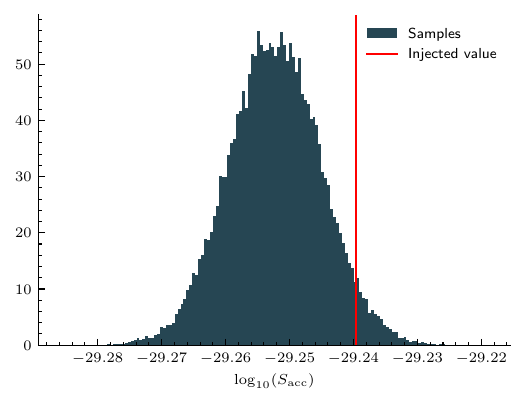}
	\includegraphics[width=\linewidth]{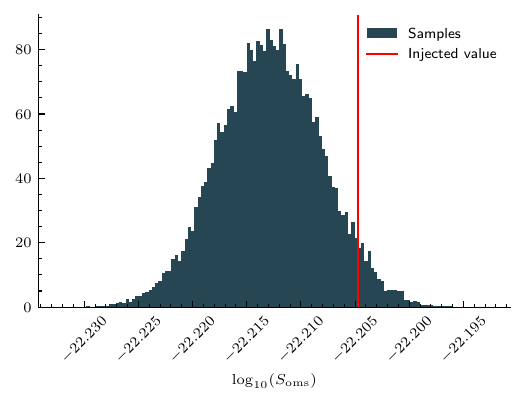}
	\caption{Conditional posterior distribution at the iteration 10
		of the acceleration noise (\textit{Up}); the optical metrology noise (\textit{Down}).
		The distribution of the noise parameters was obtained by fixing the GW sources
		identified at the 10th iteration.}
	\label{fig:noise-posteriors}
\end{figure}

We have sampled the noise parameters using uniform priors for
\(\lg(S_{\text{acc}}) \equiv \log_{10}(S_{\text{acc}})\) in
\(\mathcal{U}[\lg(\num{1.440e-31}), \lg(\num{2.304e-28})]\),
\(\lg(S_{\text{oms}})\) in \(\mathcal{U}[\lg(\num{2.500e-24}),
	\lg(\num{4.000e-21})]\), \(\lg(A)\) as \(\mathcal{U}[-50, -30]\), \(f_1\) in
\(\mathcal{U}[0.0001, 0.01]\), \(f_2\) in \(\mathcal{U}[0.001, 0.01]\),
\(\alpha\) in \(\mathcal{U}[1.0, 10.0]\), and \(f_{\rm{knee}}\) in
\(\mathcal{U}[0.0002, 0.01]\). The changes in the noise level as a function of
iteration are shown in \cref{fig:noise}. The noise level stabilises after 8 global iterations,
and, at this stage, we refer to it as the end of burn-in of the global iteration.
We also verified the convergence of the instrumental noise by comparing the
recovered parameters with the simulation. In \cref{fig:noise-posteriors} we give posterior
distribution of the acceleration noise $S_{\rm{acc}}$ and the optical metrology noise $S_{\rm{oms}}$,
using the chain of the 10th iteration. However, we should mention some caveats associated with this figure.
(i) The noise realisation was generated using several components added together. The noise model that we use
is somewhat phenomenological, there was no values $S_{\rm{acc}}$ and $S_{\rm{oms}}$ among the parameters
used for the simulation. We have reconstructed $S_{\rm{acc}}$ and $S_{\rm{oms}}$ values (vertical lines)
based on the noise components used in the simulation and the noise behaviour (model) at high and low
frequency end of the spectrum.
(ii.) We expect that the distribution will become wider as we perform more iterations of the
global fit (due to covariance with GW sources). What we plot in \cref{fig:noise-posteriors}
can be interpreted as a conditional probability density (fixing parameters of all GW sources).
All subsequent
iterations will contribute to the sampling of the conditional posteriors of the GW sources and noise.
We need a large number of the global iterations to fully explore the noise-sources correlation.


\section{Discussion}\label{sec:discussion}

In this section, we discuss what we did not explore and the path to future
improvements. First, we did not explore the verification binaries discussed in
\cref{sec:data_desc}. For those binaries, we know their precise (for GW purposes)
position in the sky, and we have a very narrow prior on the orbital frequency.
These constraints reduce the parameters volume and allow sources to be identified
with a relatively low SNR. Although our algorithm did not have special treatment
for those sources, we have detected the brightest verification GBs. We plan to
introduce the narrow-band VGB block in the iterative module. The main
difference from a normal GB block is the width of the band (to fit in one
source), the exact location of the sky, and the narrow priors for other parameters, if
available.

Second, we did not perform a time iterative analysis, which is easier than
analysing a year of data from scratch. As first mentioned in
\cite{littenberg_prototype_2023}, the iterative time analysis of GBs works like a
simulated annealing. Uploading and analysing the data growing in volume each, say 3 months, would
allow first to detect the brightest GBs. Moreover, only a little information
about the sky position is available in short data segments, which makes the
likelihood surface much shallower without sharp features that produce
secondary maxima. An analysis of 3 months, then 6 months, and so on, works as
simulated annealing (heating the likelihood and cooling it gradually down).
Note that \cite{katz_efficient_2024} uses a similar approach in
hierarchical model building. In addition, we deal gradually with incoming MBHB
mergers. The first three months of data will be analysed in exactly the same
manner as is described in the paper. We will use the posteriors obtained at
the end of the analysis as proposals and seeding points for the analysis of 6
months of data as described in \cite{korsakova_neural_2024}. In addition, we
also intend to use a more informative Galactic prior (amplitude and sky) as
suggested in \cite{korsakova_neural_2024}. Of course, we must repeat the search
for new GBs (or see if one GB source splits into two overlapped GBs) and MBHBs.
The search for GBs will consider already found sources by searching in part of
the parameter space defined by the posterior of the three-month analysis. In
the search for MBHBs, we will take into account any premerger of MBHBs
identified earlier in a similar fashion (restricting the parameter range);
otherwise, we will search for MBHBs in sequential two-week-long data segments.

Third, now that we know that the current implementation works, we are in the
process of optimising the sampling. We work on the alternative fast likelihood
computation for GBs \cite{marsat_unkown_2025}. We implement split-sampling in
``fast'' and ``slow'' parameters, where ``fast'' are usually extrinsic
parameters, which do not require recomputing the waveform or allow efficient
marginalisation (analytically or numerically). We also use broadly reweighting
techniques for sampling \cite{vallisneri_rapid_2024}.
Improvement in efficiency will allow us to use a fully Bayesian
framework using the product-space approach
\cite{hee_bayesian_2015,carlin_bayesian_1995} for multi-source (transdimensional) exploration.
Those improvements are in
the testing phase and will be applied to ``Sangria HM'' which is the simulated
data where we replace MBHB GWs with another model containing higher order modes
in addition to the dominant \((2, \pm2)\), but keeping everything else (including
parameters) the same.

\section*{Acknowledgements}
S.B.\ acknowledges funding from the French National Research
Agency (grant ANR21-CE31-0026 (project MBH\_waves) and ANR-18-CE31-0015), the European
Union’s Horizon 2020 research and innovation program under the Marie Skłodowska-Curie
grant agreement No. 101007855. We acknowledge support from the CNES for the exploration
of LISA science. The authors thank CNES for providing computational facilities.
S.D.\ acknowledges financial support from CNES.
\vfill
\bibliography{references}
\end{document}